\documentclass[10pt, a4paper]{article}
\usepackage{INTERSPEECH2016}
\usepackage{graphicx}
\usepackage{amssymb,amsmath,bm}
\usepackage{textcomp}
\usepackage{wrapfig}
\usepackage{lipsum}
\usepackage{dblfloatfix}
\usepackage{fixltx2e}
\usepackage{nameref}
\usepackage{hyperref}
\usepackage{graphicx}
\usepackage{varioref}
\usepackage{booktabs}  
\usepackage{siunitx}
\usepackage{numprint}
\usepackage{color}
\usepackage{textcomp}
\usepackage{geometry}
\usepackage{booktabs}  
\usepackage{lipsum}

\title{Comparative Analysis of SpatialHadoop and GeoSpark for Geospatial Big Data Analytics}
\makeatletter
\def\name#1{\gdef\@name{#1\\}}
\makeatother \name{{\em Rakesh K. Lenka\textsuperscript{1}, Rabindra K. Barik\textsuperscript{2}, Noopur Gupta\textsuperscript{1},Syed Mohd Ali\textsuperscript{1}, Amiya Rath\textsuperscript{3}, Harishchandra Dubey\textsuperscript{4}~\thanks{\textcolor{blue}{This material is presented to ensure timely dissemination of scholarly and technical work. Copyright and all rights therein are retained by the authors or by the respective copyright holders. The original citation of this paper is:
R.K. Lenka, R.B. Barik, N. Gupta, S. M. Ali, A. Rath, H. Dubey, "Comparative Analysis of SpatialHadoop and GeoSpark for Geospatial Big Data Analytics", India.}}}}
\address{\textsuperscript{1}Department of CSE, IIIT Bhubaneswar,India\\
\textsuperscript{2} School of Computer Applications, KIIT University\\
\textsuperscript{3} Department of CSE and IT, VSSUT Burla,India\\
\textsuperscript{4}The University of Texas at Dallas, Richardson, USA\\
{Email: rakeshkumar@iiit-bh.ac.in; rabindra.mnnit@gmail.com;noopur2827@gmail.com;}\\ {syedmohdali121@gmail.com; amiyaamiya@rediffmail.com; harishchandra.dubey@utdallas.edu}
}
\begin{document}
%
\maketitle
\begin{abstract}
In this digitalised world where every information
is stored, the data a are growing exponentially. It is estimated
that data are doubles itself every two years. Geospatial data are
one of the prime contributors to the big data scenario. There
are numerous tools of the big data analytics. But not all the
big data analytics tools are capabilities to handle geospatial big
data. In the present paper, it has been discussed about the
recent two popular open source geospatial big data analytical
tools i.e. SpatialHadoop and GeoSpark which can be used for
analysis and process the geospatial big data in efficient manner.
It has compared the architectural view of SpatialHadoop and
GeoSpark. Through the architectural comparison, it has also
summarised the merits and demerits of these tools according the
execution times and volume of the data which has been used.

\textbf{Index Terms-}Big Data; Geospatial big data; GIS; Spatial-
	Hadoop; GeoSpark.
\end{abstract}
%
\section{Introduction}
Data is exploding at an alarming rate due to growth in
various areas of business, scientific studies, academics, social
media and digitalization of all the offline records. Due to
this huge volume of data; it is difficult to perform knowledge
discovery and decision making in efficient time. According to
the International Data Corporation (IDC) report, it has been
predicted that digital data could increase by 40 times from
2012 to 2020, and it is of utmost importance that we develop
tools to handle such large amount of evolving data~\cite{b1_kune2016anatomy}. Big
Data is characterized by 5Vs, i.e. huge Volume, high Velocity,
high Variety, low Veracity and high Value [2]. Traditional
data warehousing is based on predetermined analytics over
the abstracted data but big data work on non-predetermined
analytics. Traditional technologies fail to organize and query
data like sensor data, location data, clickstream logs etc. But
we require to handle these type of data in today’s date. 

Fortunately, we have some emerging technologies like map
reduce, distributed computing etc. to tackle with this big data.
Open source frameworks like Apache Hadoop and Apache
Spark are using these technologies. These open source technologies
enable us to perform computing over very large
datasets. Hadoop is reliable, scalable and comes with software library which helps in performing distributed storage and
processing[1]. Apache Spark is fast and allows us to do batch
as well as real-time processing, it provides high level APIs in
JAVA, Scala, Python and R. High level APIs are easy to use,
require less setup but is less flexible [3].

One of the emerging field in big data is Geographical
Information System (GIS). According to NCGIA, “GIS is a
system of hardware and software which facilitate the management,
manipulation, analysing, modelling, representation
and display of georeferenced data to solve complex problem
regarding planning and management of resources by using of
so many various types of open source GIS software [4] [6].
Since geographic data are in such huge volume and velocity it
qualifies itself as a Big Data problem. The above-mentioned
technologies are required to store and process geospatial data
in efficient way.
\section{GEOSPATIAL BIG DATA}
\label{sec:s2}
Geospatial Data is the data related to geographical location,
it is usually stored as coordinates and topology and it is used
for mapping [5]. Geospatial Big Data is not the new problem,
in 2007 itself it crossed the storage capacity due to exponential
increase in data production. Considering the velocity of data,
Remote Sensing data archives of EOSDIS are growing at the
rate of 4TB daily. The data flow offers to the users around the
world every day would be about 20 TB, which means more
than 630 million data files. Every unit time, the observing
data gathered from nearly 100 active missions of NASA would
be about 1.73GB [7]. NASA satellite data archives exceeded
500TB and is still growing [8]. Sources of these data are
satellite remote sensing, digital camera, sensor networks, radar,
LIDAR etc. To model and simulate these huge volume of
geospatially enabled content, we require high performance
computing or cloud computing environment more than ever
[9]. There are three kinds of geospatial data [10].
\begin{itemize}
\item Raster Data: It contains images taken by digital cameras,
satellite etc. It is a regular grid of pixels of fixed size and
is best suited for continuous data.

\item Vector Data: They are build using points, lines and
polygons. It is best suited for categorical or discrete data.
The map data belongs to this form. Ex: Sea, forest, land etc.

\item Graph Data: It appears in the form of city maps containing
roads and landmark. Roads are represented as edges and landmarks or intersection as node.
\end{itemize}
Earth consists of many complex features and GIS helps
us to keep track of these features in a simplified manner.
The more complex the feature, more storage required, more
computational complex queries. GIS is a tool that helps to
manage, analyse, and display spatial information on a computer.
GIS helps us to skip traditional process of drawing
map, which can be time consuming and expensive. GIS
combines map making with Database management system.
GIS stores graphical operation of real world features and
information about the real world features. These features are
linked geographically to some kind of coordinate system. The
data can be about people, population, income, education, land,
soil types, water resources, storm drains, electric power line,
etc. Data that are in digital form can be directly uploaded
into GIS whereas maps can be scanned or converted into
digital information. On a single map one can show different
kinds of data through GIS, which helps in analysing and
understanding patterns and relationships. GIS can be used
as a tool to explore other planets. For example, the Global
Biodiversity Information Facility (GBIF) has collected more than 400 million species occurrence records and most of them
comes with (latitude, longitude) pair. To find bio diversity
patterns it is important to map all of these species. This
mapping and querying exceeds the capability of the traditional
system and hence qualifying for the domain of big data. Now a
days, Fog computing comes into the picture for geospatial big
data analytics in health geoinformatics, smartwatch technology
and location based recommender system [11] [12] [13] [14]
[15].
\begin{figure}[!t]
\includegraphics[width=240bp]{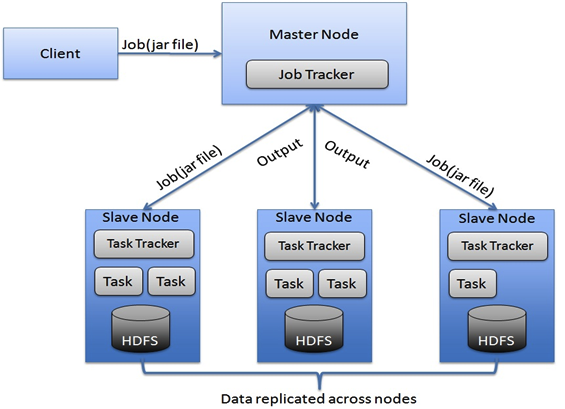}
\caption{Architecture of Hadoop Ecosystem.}
\label{fig1}
\end{figure}
\section{SPATIAL-HADOOP}
Apache Hadoop is an open source framework written in
JAVA to perform distributed storage and computing of very
large dataset [3]. Traditional approaches used powerful computer
to process these huge data but those were not scalable,
so at some point of time they were not able to cope with
continuous growth of data. Hadoop solved this problem with
the help of distributed computing as it is scalable. Hadoop
breaks down the big data into smaller parts hence able to
achieve distributed storage. It also breaks down the processing
into chunks and every node performs its subtask, ultimately
result of all the nodes are combined to give the final result
which is returned to the application, in this way parallel and
distributed processing is achieved. Figure~\ref{fig1} shows the general
architecture of Hadoop Ecosystem.

Hadoop has its own storage area, processing area known
as Map Reduce and Hadoop Distributed File System (HDFS)
[3]. Map Reduce divides the large dataset into independent
small data which will be processed in parallel where as HDFS spreads multiple copy of data into multiple machine
which gives us reliability which has been shown in Figure 2.
This task is done automatically by Hadoop Ecosystem hence
programmer need not worry about it, which means we can
write scale free programmes.

We initially had Hadoop-GIS for this which was built on
Hive, but this thing became too difficult as we were unable
to do traditional map reduce job. Whereas SpatialHadoop
is directly integrated into in the Hadoop making it more
efficient than Hadoop-GIS [16]. SpatialHadoop provided tools
for the developers and researchers to implement new spatial
operations in the system efficiently. We need to use hadoop
but unfortunately hadoop is not able to process spatial data
efficiently. For example: Range query is better performed by
SpatialHadoop compared to range query perform by hadoop as
it takes less time on SpatialHadoop. But this type of facilities
was not present in Hadoop-GIS and their counter parts hence
they were very limited to the functionalities they came with.
For the visualization purpose of geospatial big data, we have
a mapreduce based framework that is built on top of Hadoop
and is known as Hadoopviz.

Hadoopviz is one of the open source tool and can produce
giga-pixel images for spatial big data. It is an important tool
to learn about geospatial data visualization which helps in
detecting interesting patterns and relationships in the given
data easily and quickly. Hadoopviz is extensible and this
makes it more industry-friendly. It provides a large variety
of data visualizations and it uses a 3-phase technique called
partition-plot-merge. Hadoopviz can visualize in two forms,
they are single level visualization and multilevel visualization.
In single level visualization, it produces an image that shows
all information in single level whereas in multilevel visualization
the information is shown through multiple levels and
produces multilevel gigapixel images [17].
\begin{figure}[!t]
	\includegraphics[width=240bp]{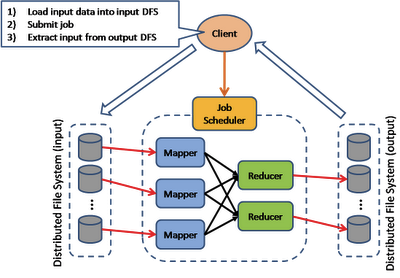}
	\caption{Architecture of MapReduce.}
	\label{fig2}
\end{figure}
\section{GEOSPARK}
\label{sec:s4}
\begin{figure}[!t]
	\includegraphics[width=240bp]{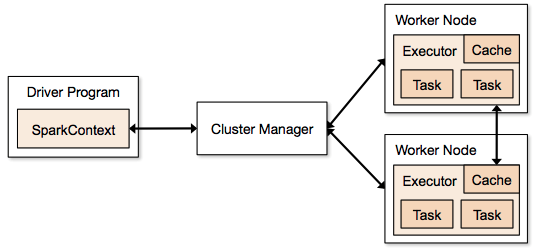}
	\caption{Cluster mode overview.}
	\label{fig3}
\end{figure}

Spark is the hottest tool to analyse and work on Big Data.
Now-a-days most of the companies are using Apache Spark
as it provides batch processing as well as real time processing
and it is 100 times faster than Hadoop. Spark is written in
Scala and performs in-memory computation. Some high level
tools supported by Apache Spark are Spark SQL, Spark MLlib,
GraphX, Hive and Spark streaming [3]. When it combines
with Hive it is known as SHARK, and SHARK is 100 times
faster than Hive. Machine learning algorithms are best suited
in this framework and it is an interesting feature of Spark.
It not only does faster batch processing but also does Realtime
processing, interactive data analysis which in turn makes
faster decision making. It also supports iterative algorithm; this
is not easily done in Hadoop. Apache Storm is also a good
contender for the real time processing but we use Spark over
it because of the following reasons:
\begin{itemize}
\item It does not support integration with Hadoop
\item It only supports real time processing
\end{itemize}
Apache Spark provides us both reliability and speed by giving
all the features which Storm could not. Spark has two type of programs i.e. Driver Program which
run on Master andWorker Program which runs on Slave. Spark
Context is responsible for the connection to the Cluster Manager
which in turn allocates the resources on the slave(worker)
nodes which has been shown in Figure~\ref{fig3}.
Spark is extremely flexible as not only it has its own cluster
manager but it connects to other sources like Apache Memos and Hadoop (HDFS). It has integrated with Hadoop as they
both belong to the same family so if you are running Hadoop
and you want to switch over to Spark then you can run Spark
on the top of the Hadoop hence saving all the data migration
time from HDFS to Spark. It can also combine with Apache
Cassandra, OpenStack Swift, Amazon S3.

As we are witnessing that Geospatial data is becoming a
Big Data problem and people were using Hadoop to tackle
geospatial problem. Since Spark is so much faster than Hadoop
people tried to use Spark instead of Hadoop to solve various
type of analysis of geospatial data, but unfortunately Spark
had no support for Spatial data and operation. GeoSpark tries
to solve this problem by extending the core of Apache Spark
and providing tools to tackle geospatial data [18].
\section{OBJECTIVE OF THE PRESENT STUDY}
\label{sec:s5}
In the preent paper, it has discussed about different frameworks
like Hadoop, spark and the frameworks which are built
on top these Hadoop and Spark like Hadoop-GIS, HadoopViz,
SpatialHadoop and GeoSpark. All these frameworks are used
for geospatial big data analytics. The objective the present
study is to make everyone familiar about these tools, about
their architecture and working as they are specially built for
geospatial data and the people working in the field of geospatial
data may need these tools. Hadoop-GIS built to handle
geospatial big data but was unable to perform mapreduce job.
SpatialHadoop is built on the top of hadoop and can perform
different geospatial operations on the given geospatial big
data. HadoopViz, this was also built on top of Hadoop and
is used for visualization of geospatial big data in gigapixels.
GeoSpark is built on the top of Spark and can perform
geospatial operations on given geospatial data but it is faster
than SpatialHadoop. This study makes it clear that which tool
is required for differnt tasks related to geospatial big data.
\section{ARCHITECTURE COMPARISON}
\label{sec:s6}
Since It has been dealt with SpatialHadoop and GeoSpark
for geospatial big data analytics, it is important to know the
difference between their architecture and working.
\subsection{Architectural view of SpatialHadoop}
SpatialHadoop is a full fledged MapReduce framework
with native support for geospatial data and it overcomes the
limitation of Hadoop-GIS. The architecture of SpatialHadoop
consists of the following four layers[19].
\begin{itemize}
\item Language Layer: The language used in SpatialHadoop is
pigeon which is an extension of pig. It is because of this
language only that SpatialHadoop has the ability to add
geospatial datatypes, functions and operations.

\item Operation Layer: The different type of geospatial operations
can be performed by SpatialHadoop like range
query, KNN, geospatial join etc. We can add more
operations in this layer like KNN join, RNN etc.

\item MapReduce Layer: This layer of SpatialHadoop runs
mapreduce program. The input files supported by SpatialHadoop
are geospatially indexed. It is better that traditional Hadoop as it adds two main components as
follows:
\begin{enumerate}
	\item Geospatial File Splitter: It is an extension of a file
splitter that is used in Hadoop system.
\item Geospatial Record Reader: It reads the slitted file
originating from input files and for efficient processing
it uses local index.
\end{enumerate}

\item Storage: In Hadoop the input files are non indexed heap
files, but SpatialHadoop makes geospatial index structure
within the HDFS. Due to this, it has two-level index structure
of global and local indexing. Global index divides
the data across all the nodes while the local indexes with
the node. Architectural view of SpatialHadoop has been
illustrated in Figure~\ref{fig4}. 
\end{itemize}
\begin{figure}[!t]
	\includegraphics[width=240bp]{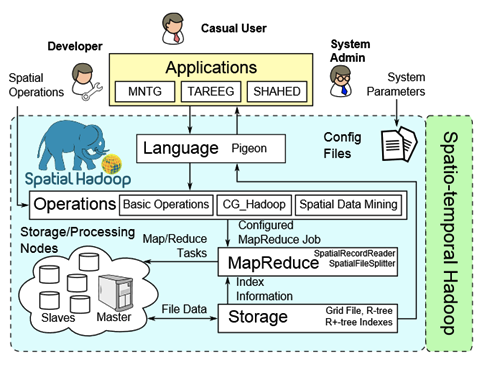}
	\caption{Architectural view of SpatialHadoop.}
	\label{fig4}
\end{figure}
\subsection{Architectural view of GeoSpark}
To handle large volume of geospatial data , it has used
GeoSpark as it is an in-memory cluster computing system.
It is an extension of Apache Spark which supports geospatial
datatypes, indexes and operations which has been illustrated
in Figure~\ref{fig5}.
\begin{figure}[!t]
	\includegraphics[width=240bp]{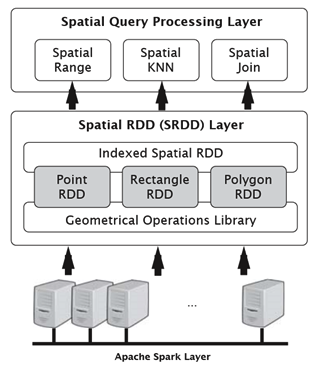}
	\caption{Architectural view of GeoSpark.}
	\label{fig4}
\end{figure}
The architecture of GeoSpark consists of the following three
layers:
\begin{itemize}
\item  Apache Spark Layer: It consists of all the components
present in Spark. It performs loading and querying data.
\item Geospatial Resilient Distributed Dataset Layer: This layer
extends the Spark. There are three types of RDD in this
layer i.e. Point, Rectangle and Polygon RDD. It contains
geometrical operations library for every RDD.
\item Geospatial Query Processing Layer: It is used to perform
different types of geospatial queries
\end{itemize}
\section{RESULT and DISCUSSION}
\label{sec:s7}
In the present research paper, it has discussed the architectural
view of both the open source frameworks. Both of these
tools are powerful and handy to use and can efficiently handle
geospatial big data analytics. It has the capability to add more
functionalities and operations in each of these tools as per
the requirements. Figure 6 and Figure 7 describe the runtime
analysis between SpatialHaddop and GeoSpark according to
the cluster size. From the graph, it has been clearly shown that
GeoSpark has the edge over the SpatialHadoop for geospatial
big data analytics when the cluster size compared with time
span for geospatial big data processing.
\begin{figure}[!t]
	\includegraphics[width=240bp]{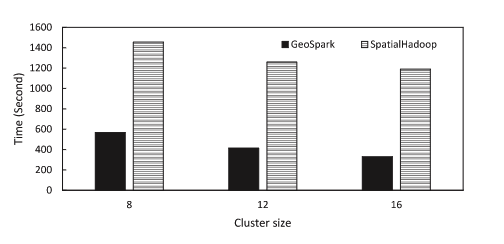}
	\caption{Run time analysis between SpatialHadoop and GeoSpark[18].}
	\label{fig6}
\end{figure}
\begin{figure}[!t]
	\includegraphics[width=240bp]{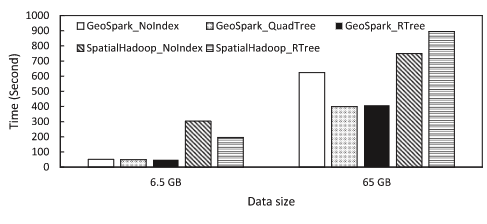}
	\caption{Run time analysis[18].}
	\label{fig7}
\end{figure}
Since no tool is perfect and have merits and demerits. Table
1 represents the merits and demerits of SpatialHadoop and
GeoSpark.
%
\begin{table*}[!t]
\centering
\caption{GEOSPATIAL BIG DATA ANALYTICS TOOLS.}
\begin{tabular}{|c |c| c|}
\hline
\textbf{Tools} & \textbf{Merit}& \textbf{Demrit} \\ \hline
SpatialHadoop & 1-Uses MapReduce framework,& Extremely slow \\ 

                        &  2-Basic functionalities of Hadoop like reliability,& when compared to  \\
                        & availability, scalability etc. &   GeoSpark[18] \\ \hline
                        
GeoSpark &  1-Uses MapReduce framework, & 1-Limited community
support, \\ 
 & 2-Derived from Apache Spark, & 2-Only supports JAVA
and SCALA, \\
 & 3-Much faster than
SpatialHadoop & \\ \hline
\end{tabular}
\label{table1}
\end{table*}
\section{Conclusions}
In this paper, it has discussed the two important geospatial
big data analytical tools i.e. SpatialHadoop and GeoSpark for
geospatial data processing. It has also described the certain
architectural comparative analysis between these two open
source tools which handle geospatial big data efficiently which
build on the top of Hadoop and Spark. Both GeoSpark and
Spatial-Hadoop are efficient in handling geospatial data but
GeoSpark has edge over Spatial-Hadoop as it is much faster for
realtime processing of geospatial big data. In future studies, it
has been planned to implement in disaster mitigation and management
as well as geospatial health information infrastructure
management for better predictive analytics.
\newpage
\eightpt
\nocite{*}
\bibliographystyle{IEEEbib}
\bibliography{sh}

\end{document}